\documentclass{jkas}

\def\year{2020} 
\def\month{April} 
\def\volume{00} 
\def\issue{0} 
\def\beginpage{1} 
\def\endpage{99} 
\def\received{---} 
\def\accepted{---} 
\date{Received \received; accepted \accepted}

\title{
Simulations of Torus Reverberation Mapping Experiments with SPHEREx
}

\author[1]{Minjin Kim}
\author[2,3]{Woong-Seob Jeong}
\author[2,3]{Yujin Yang}
\author[1]{Jiwon Son}
\author[4,5]{Luis C. Ho}
\author[6,7]{Jong-Hak Woo}
\author[6,7]{Myungshin Im}
\author[2,3]{Woowon Byun}

\affil[1]{Department of Astronomy and Atmospheric Sciences, 
College of Natural 
Sciences, Kyungpook National University, Daegu 41566, Korea; 
\email{mkim.astro@gmail.com}}
\affil[2]{Korea Astronomy and Space Science Institute, 
Daejeon 34055, Korea}
\affil[3]{University of Science and Technology, Daejeon
34113, Korea}
\affil[4]{Kavli Institute for Astronomy and Astrophysics, Peking University, 
Beijing 100871, China}
\affil[5]{Department of Astronomy, School of Physics, Peking University, Beijing 
100871, China}
\affil[6]{Astronomy Program, Department of Physics and Astronomy, Seoul National University, Seoul 08826, Korea}
\affil[7]{SNU Astronomy Research Center, Seoul National University, Seoul 08826, Korea}

\def\corrauthor{
M. Kim
}

\def\runningauthor{
Author Kim et al.
}

\def\runningtitle{
Torus RM with SPHEREx
}

\def\keywords{
black hole physics --- 
galaxies: active --- galaxies: Seyfert --- quasars: 
general --- infrared: galaxies --- dust --- surveys
}

\def\abstracttext{
Reverberation mapping (RM) is an efficient method to investigate the
physical sizes of the broad line region (BLR) and dusty torus in an active 
galactic nucleus (AGN). The Spectro-Photometer for the History of the Universe, 
Epoch of Reionization and Ices Explorer (SPHEREx) mission will provide 
multi-epoch spectroscopic data at optical and near-infrared wavelengths.
These data can be used for RM experiments for bright AGNs. We present 
results of a feasibility test using SPHEREx data in the SPHEREx deep regions 
for the torus RM measurements. We investigate the physical 
properties of bright AGNs in the SPHEREx deep field. Based on this 
information, we compute the efficiency of detecting torus time lags in 
simulated light curves.  We demonstrate that, in combination with the 
complementary optical 
data with a depth of $\sim20$ mag in $B-$band, lags of $\le 750$ days for 
tori can be measured for more than $\sim200$ bright AGNs. 
If high signal-to-noise ratio photometric data with a depth 
of $\sim21-22$ mag are available, RM measurements can be applied for up to 
$\sim$900 objects. When complemented by well-designed early optical 
observations, SPHEREx can provide a unique dataset for studies of the physical 
properties of dusty tori in bright AGNs.    
}

\begin{document}
\jkashead 


\section{Introduction}
An active galactic nucleus (AGN) is believed to originate from 
the strong thermal/non-thermal emission generated from the accretion disk around 
a supermassive black hole (SMBH) at the center of galaxies. 
Observational and theoretical studies suggested that SMBHs dominantly gain 
their mass from accreted gas during the bright AGN phase [Quasi-stellar Objects
(QSO) phase]. 
Hence, knowledge of the AGN is essential for understanding the formation and 
evolution of an SMBH in the Universe (e.g., \citealt{yu_2002}).
According to the AGN unification model (\citealt{antonucci_1993, urry_1995}), 
the central region of an AGN is composed of the accretion disk closely 
associated with an SMBH, broad line region (BLR), dusty torus, 
and narrow line region (NLR). 
Therefore, studying physical properties
of these sub-components are of great importance to understand the 
innermost structure of AGNs (e.g., \citealt{blandford_1982, lopez_2016};
Gravity Collaboration 2020b). 

Because the physical sizes of the central structures, except the NLR, is 
relatively small ($\le$ a few pc), exploring their structures directly with
high-resolution images has been limited to nearby AGNs (e.g., 
\citealt{jaffe_2004}; Gravity Collaboration 2020a; 2020b).  
In this context, the reverberation mapping (RM) method has been
widely used to investigate the physical size of central structures
(\citealt{blandford_1982}). 
The UV/optical featureless continuum originates from the accretion disk, 
and the broad emission line arises from the photoionization of the gas in 
the BLR, which is ionized by the UV photon from the accretion disk.
Therefore, the flux of the broad emission line varies in response to the 
variation in the optical continuum with a time delay (i.e. time lag). This time 
lag between two components is a good indicator of the physical distance between 
the accretion disk and BLR. 

RM method has been widely applied to measure the size of the BLR in AGNs. 
Using knowledge of the line width of the broad emission line along with RM
can provide estimates of the mass of the SMBH by the virial method. 
Interestingly, the size of the BLR is reported to be strongly correlated
with the luminosity of AGN by the size-luminosity relation
(e.g., \citealt{kaspi_2000, bentz_2013, du_2016}).
Using this correlation, one can estimate the BH mass even using a single-epoch
spectrum with an uncertainty of $\sim0.3-0.4$ dex (\citealt{onken_2004,
woo_2010, ho_2014}). 
However, RM experiments require a large amount of observation time. 
Hence, BLR time lags have been successfully measured for $\sim$ a hundred 
of AGNs so far (e.g., \citealt{peterson_2004, bentz_2009, barth_2015, kim_2019, 
rakshit_2019}). While the on-going RM projects with 
multi-object spectrograph can increase the sample size significantly, they are 
still insufficient to investigate the BLR properties for a diverse range of 
physical properties of AGNs (e.g., Eddington ratio and black hole 
mass; \citealt{king_2015, shen_2015, grier_2017, du_2018, shen_2019}).
The BLR RM with SPHEREx will be discussed elsewhere.

\begin{figure}[t]
\centering
\includegraphics[width=80mm]{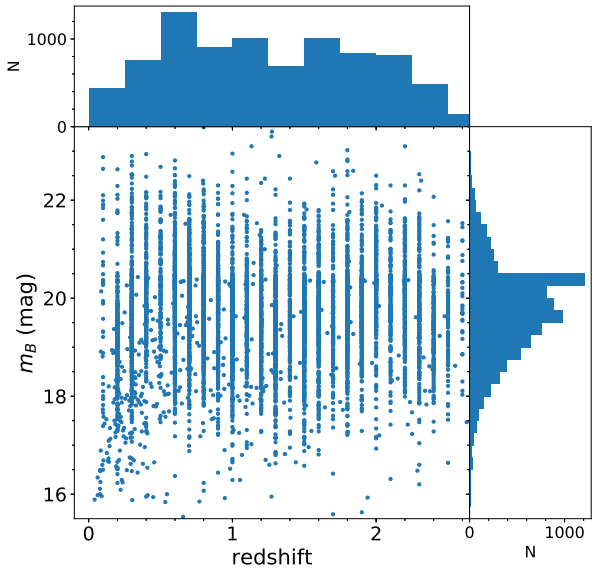}
\caption{Distributions of redshift and $B-$band apparent magnitude ($m_B$) of
target QSOs in the SPHEREx deep regions.\label{fig:fig1}}
\end{figure}
\vspace{0.5cm}
 
The RM method can be also applied to estimate 
the size of the dusty torus. The dust in the torus is heated by the UV 
emission from the accretion disk and radiates thermal energy mostly in the 
infrared (IR). Similar to observation in  RM experiment with BLR, in this case 
too, the time lag between the IR and UV emissions provides with a constraint on 
the sublimation radius. 
The torus RM measurements are even more limited to a few dozens of AGNs 
because intensive IR monitoring with ground-based telescopes is a challenge 
(e.g., \citealt{clavel_1989, glass_2004, koshida_2014}). Recently, 
the IR survey data from Wide-field Infrared Survey Explorer 
(WISE; \citealt{wright_2010}) in combination with the optical photometric data 
from various transient surveys were successfully used in RM experiments for the
torus (\citealt{lyu_2019}). However, the WISE multi-epoch data are 
obtained with a cadence of 6 months, resulting in relatively large 
uncertainties in the lag measurements for the individual target. 

The Spectro-Photometer for the History of the Universe, Epoch of Reionization 
and Ices Explorer (SPHEREx) mission will conduct an all-sky spectral survey 
to cover a spectral range of $0.75-4.5 \mu{\rm m}$ with a $5\sigma$ depth
of {\rm 19 mag\footnote{This depth can be achieved with a single visit.}} 
in each spectral bin (\citealt{dore_2016, dore_2018}). 
The mission will be launched in late 2023 and perform the all-sky mapping 
four times during 2 years. It will employ linear variable filters 
to obtain the spectral imaging data with a spectral resolution of 
$\sim40$. Because of the wide field of view of detectors, SPHEREx will cover the 
deep regions ($\sim 200$ ${\rm deg}^2$) around the equatorial poles more than 
100 times in 2 years. The multi-epoch dataset in the deep survey fields
will allow us to conduct RM experiments for both the BLR and torus, owing 
to the optical/NIR wavelength coverage and spectral capability. 
However, due to the relatively short baseline ($\sim2$ years) of the SPHEREx mission,
time lags larger than 2 years are not effectively detected solely with 
SPHEREx data. Therefore, early complementary optical monitoring data is 
strongly necessary to increase the detection efficiency of bright AGNs 
(\citealt{shen_2015}). 

The goals of this work are (1) to explore the feasibility of {\it torus} RM 
measurements with SPHEREx dataset and (2) to find the optimal way to obtain 
complementary optical data. 
The paper is organized as follows. 
In section 2, we report a search for bright AGNs 
(i.e., QSOs) in the SPHEREx deep fields. In section 3,
we describe our methods to generate the simulated light curves. In section 4,
we present the results of RM measurements from the simulated light curves. 
In section 5, we discuss the best way to maximize the efficiency of the RM 
experiments. Throughout this paper, we have adopted the following 
cosmological parameters: 
$H_0 = 100 h = 67.8$ km s$^{-1}$ Mpc$^{-1}$, $\Omega_m = 0.308$, and 
$\Omega_{\Lambda} = 0.692$ (\citealt{planck_2016}).

\begin{figure}[t]
\centering
\includegraphics[width=80mm]{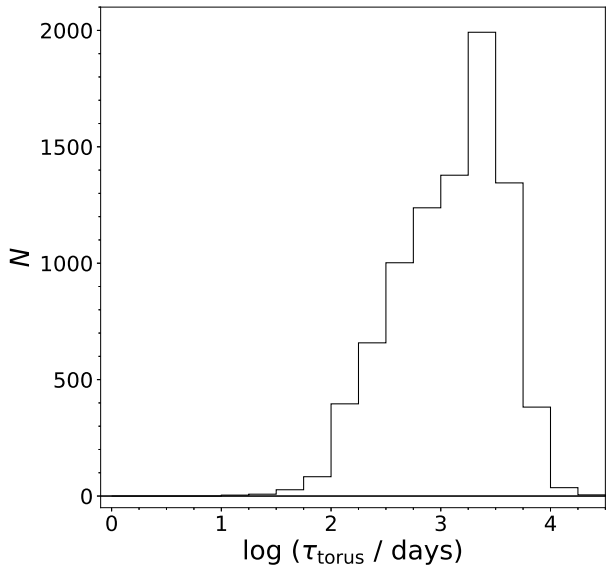}
\caption{Predicted distributions of the observed time lags for the torus.
\label{fig:fig2}}
\end{figure}
\vspace{1.5cm}

\section{Sample Characteristics}
It is vital to investigate the physical properties of target QSOs in SPHEREx 
deep fields around north and south ecliptic poles from the two perspectives 
of testing the feasibility of the RM experiments and identifying
the best strategy for obtaining the complementary optical imaging observation. 
In this context, the most essential properties are the brightness of the
potential targets and expected time lags for the BLR and torus.  
To search for the targets, we assume that the north and south deep 
fields are centered on R. A.= 17:55:24 Decl. = 66:37:32 and R. A. = 4:44:00 
Decl. = $-$53:20:00, respectively. Each field covers 100 deg$^2$ with a radius 
of 5.64 degree. Although the positions of the deep fields are not yet
determined, the statistical properties of target QSOs should not be
sensitive to the exact positions, considering the wide area of the fields.

We use the Million Quasars (MILLIQUAS) Catalog, Version 6.5 
(\citealt{flesch_2019}) to search for potential targets for the 
RM studies. We find 5867 and 3674 QSOs in the north and 
south fields, respectively. Because those regions had not been covered by the
large spectroscopic surveys such as Sloan Digital Sky Survey (SDSS), 
2dF and 6dF galaxy surveys, the majority of the sample comprises the QSO 
{\it candidates} selected from multiwavelength photometric data 
(e.g., WISE; \citealt{secrest_2015}),
which are yet to be confirmed using spectral data. The photometric redshifts 
of the QSO candidates were estimated based on optical and mid-infrared (MIR) 
colors (\citealt{flesch_2015}). The uncertainty of these redshifts is 
approximately 
50\%\footnote{https://heasarc.gsfc.nasa.gov/W3Browse/all/milliquas.html}.
Figure 1 shows the distributions of the redshift and 
apparent magnitude in the $B-$band ($m_B$). 
Because the completeness of the MILLIQUAS catalog is unknown, we 
independently estimate the expected number of QSOs in the deep fields. Using 
the number counts of QSOs for $z<2.1$ (\citealt{richards_2005}), 
at $m_B<20$ mag, $\sim3500$ QSOs are expected to be present in a field of 
200 deg$^2$, while we find $\sim 5600$ objects from MILLIQUAS catalog. 
Taking into account of predicted type 1 fraction ($\sim50$\%; see below) in 
the sample, two values appear to be broadly consistent each other.

In order to estimate expected time lags, we adopt the 
size-luminosity relation reported by \citet{koshida_2014} for the torus.
For the sublimation radius for the torus, 
the relation is expressed as $R_{\rm sub}= -2.11-0.2\times M_V$, where 
$R_{\rm sub}$ is
the dust sublimation radius measured from the $K-$band, given in the unit
of light days. We convert $m_B$ to $L_{5100}$ 
(the monochromatic luminosity at 5100\AA) and $M_V$ using the 
optical/IR spectral template of QSOs (\citealt{hickox_2017}).  
Finally, we account for the cosmological time delay ($1+z$) for 
estimating the observed time lags $\tau_{\rm torus}$.

\begin{figure*}[h]
\centering
\includegraphics[width=155mm]{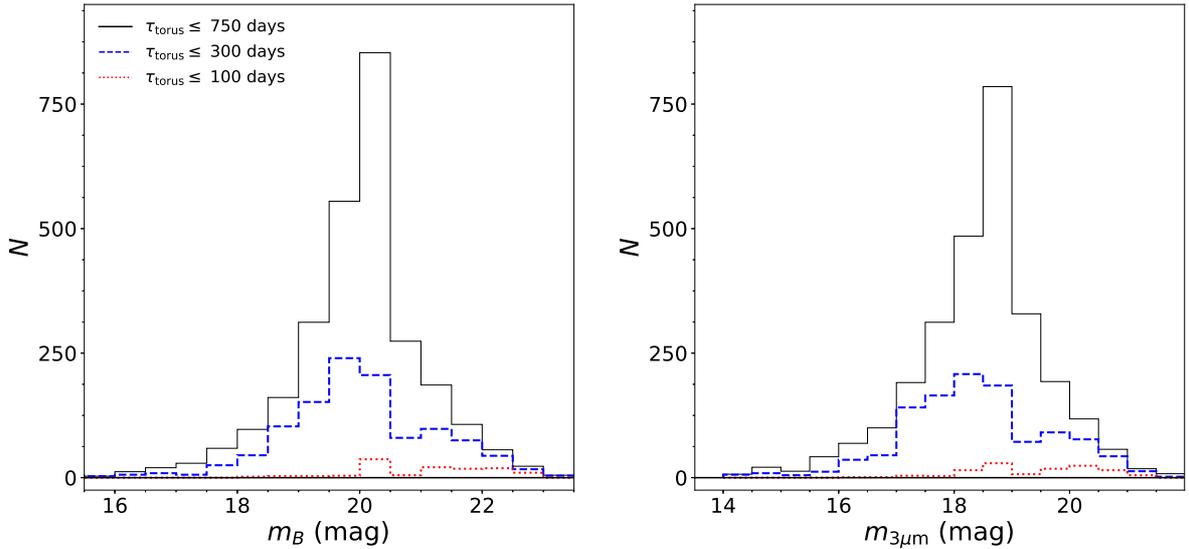}
\caption{Distributions of the apparent magnitudes at $B-$band (left) 
and $3\mu {\rm m}$ (right) of the primary targets ($\tau_{\rm torus} \le 750$ 
days. The IR brightness is computed using the QSO template from 
\citet{hickox_2017}.
\label{fig:fig3}}
\end{figure*}

The predicted distribution of $\tau_{\rm torus}$ is presented in Figure 2. 
The uncertainty in the estimations ($\approx 0.25$ dex) is mainly introduced 
by the lack of spectroscopic redshift and the intrinsic scatter 
($\approx0.13$ dex) in the size-luminosity relations (\citealt{koshida_2014}). 
Considering 2-year mission of 
SPHEREx, time lags over 750 days may not be detectable even with 
earlier optical imaging data obtained for 2 years prior to 
the SPHEREx mission. Using the simulation described in \S{3}, we also confirm 
that the time lags above 2 years are barely detected. 
Therefore, we consider QSOs with $\tau_{\rm torus}$
smaller than 750 days ($\sim2$ years) as the primary targets. 
Throughout the paper, only the primary targets are discussed. 
Overall, we found 2785 (1129) QSOs with $\tau_{\rm torus} \le 750$ (300) 
days. 

The fraction of type 1 in the IR-selected AGNs strongly depends on
the bolometric luminosity (e.g., \citealt{assef_2015}). 
Therefore we compute the bolometric luminosity ($L_{\rm Bol}$) using 
$L_{5100}$ and the bolometric conversion factor 
($L_{\rm Bol}=9.26\times L_{\rm 5100}$; \citealt{richards_2006}). 
It yields $L_{\rm Bol}$ ranges from $10^{44.5-46}$ erg s$^{-1}$ with 
a median value of $10^{45.4}$ erg s$^{-1}$ for the primary target. 
The type 1 AGN fraction is expected to vary from 0.3 to 0.7 
(equation 15 of \citealt{assef_2015}). For the sake of simplicity, 
we assume that a half of the QSO candidates are type 1.

Figure 3 shows the distributions of the brightness of the primary targets. 
We assume that the flux at $3\mu {\rm m}$ will be mainly used
for the IR RM study, which approximately corresponds to the $K-$band
in the rest frame for the median redshift of the target QSOs.
Because, for high-redshift targets, the contamination from the accretion 
disk may not be negligible even at $3\mu {\rm m}$ (\citealt{sakata_2010,
koshida_2014,honig_2014}), we will make use of the flux at longer 
wavelength. In addition, thanks to the wide spectral coverage of the 
SPHEREx dataset, we will also investigate the color variation in NIR, 
which will provide useful constraints on the structure of the torus.

The AB magnitude at $3\mu {\rm m}$ is computed from $m_B$ using the 
optical/IR QSO template of \citet{hickox_2017}. Approximately $76\%$ of the 
primary targets is expected to be 
brighter than 19 mag at 3$\mu {\rm m}$, indicating that the IR 
variability can be robustly measured by the SPHEREx survey. 
Further, 19 mag at 3$\mu {\rm m}$ approximately corresponds to 20 mag in the 
$B-$band. Therefore, the optical imaging data to trace the UV/optical
variability are necessary to reach 21-22 mag with a $5\sigma$ detection limit 
to maximize the sample size.  
In addition, the QSO targets fainter than 19 mag can also be used for 
the RM study by increasing the signal-to-noise (S/N) through temporal binning, 
or spectral binning. 
The spectral binning will be performed mainly within $\sim9-10$ spectral
bins ($3\pm0.4\mu {\rm m}$), in order to minimize the contamination from
the accretion disk and preserve the color information. 
Overall, taking into account of the type 1 fraction ($\sim 50$\%), 
we found that sufficient samples of QSOs 
($\sim 1400$) are available in the SPHEREx deep fields and can be used 
for RM measurements. These data will allow us to explore 
the central structure of QSOs comprehensively. 

\section{Simulation}
\subsection{Light Curve}
To test the feasibility of the RM experiments in the SPHEREx deep fields, 
we perform extensive simulations with the artificial light curves of AGNs. 
In general, the light curve of AGNs is well modelled with a broken 
power law (\citealt{mushotzky_2011, kasliwal_2015, caplar_2017}). In this 
model, 
the power spectrum at low frequencies is moderately flat, and that above the 
break frequency is steep. Here, we adopt the damped random walk (DRW)
model to generate the light curves, which is known to well represent the 
observed light 
curve of bright AGNs (\citealt{macleod_2012}). According to DRW, we assume that 
$\gamma$ is 2 at high frequencies ($\propto \nu^{-\gamma}$), and that 
$\gamma$ is 0 at lower frequencies.
While the break frequency ranges from 10 to 150 days, we assume 
a break frequency is 100 days in the simulation, for the sake of simplicity
(see also \citealt{woo_2019}). 

The variability amplitude is generally known to be $\sim10\%$ in the 
optical band for bright QSOs (e.g., \citealt{giveon_1999, sanchez_2018}). 
Therefore we firstly assume that the variability amplitude is 10\%.
Several studies claimed that the variability amplitude is increasing with
decreasing the luminosity of AGNs (e.g., \citealt{sanchez_2018}). Our sample
of the primary targets comprises a relatively faint QSOs 
($43 \leq \log (L_{5100}/{\rm erg s^{-1}}) \leq 45$), suggesting that 
our assumption for the variability amplitude is conservative. 
However, the amplitude in the IR is significantly smaller than that in the 
optical band (\citealt{lyu_2019}). In addition, owing to the low spatial 
resolution of SPHEREx ($\sim6$ arcsec), light contamination from the host 
galaxy is also non-negligible. Using the high resolution images of nearby QSOs 
obtained with {\it Hubble Space Telescope} (\citealt{kim_2017}), we find that 
the host brightness is comparable to the nuclear brightness on average, 
indicating that the variability amplitude can be reduced by $\sim 25$\%. 
In order to take these effects into 
account, we examine two cases (5\%\ and 10\%\ of the variability 
amplitude) for generating the light curves.
To investigate the dependence on the brightness of the targets, we 
artificially added photometric uncertainty ranging from 
1\%\ to 10\%\ of the original flux. To take into account 
the expected distributions of the observed time lags in the deep fields, 
we consider three cases (100, 300, and 700 days).
Because our goal is to demonstrate the general feasibility of the IR 
RM experiment with SPHEREx and to offer the best strategy, we simply assume 
that the variability amplitude and photometric uncertainty are same for both 
optical and IR band in this experiment.

\begin{table*}
\centering
\caption{
Observation Strategies
\label{tab:jkastable1}}
\begin{tabular}{cccccc}
\toprule
epoch & \multicolumn{2}{c}{Optical Data} & & \multicolumn{2}{c}{SPHEREx} \\
\cline{2-3} \cline{5-6}
 & Cadence1 & Cadence2 & & Cadence & Sea. Gap \\
(1) & (2) & (3) & & (4) & (5)  \\
\midrule
epoch1 (Opt2 + IR1) & 1 month & 2 weeks & & 6 days & 6 months \\
epoch2 (Opt1 + IR1) & 1 month & 1 week  & & 6 days & 6 months \\
epoch3 (Opt2 + IR2) & 1 month & 2 weeks & & 6 days & 9 months \\
epoch4 (Opt1 + IR2) & 1 month & 1 week  & & 6 days & 9 months \\
\end{tabular}
\tabnote{
Col. (1): Name of the observing strategy.
Col. (2): Cadence of the complementary optical data in the first 1.5 years.
Col. (3): Cadence of the complementary optical data in the second 2.5 years.
Col. (4): Cadence of the SPHEREx data.
Col. (5): Seasonal gap for SPHEREX data.}
\end{table*}

\subsection{Observation Strategy}
The success rate of detecting the time lags is very sensitive to 
the designed epochs and cadences of the monitoring program. 
Therefore, we use two observing sequences for each of 
the complementary optical imaging data and IR spectroscopic data 
obtained from SPHEREx. To overcome the relatively short baseline
of the SPHEREx mission ($\sim2$ years), the optical monitoring data from
the ground-based telescopes is assumed to be obtained for 4 years, 
starting 2 years 
prior to the beginning of SPHEREx mission and over the duration of the mission.
In the first scenario (called ``Opt1'' through out this paper), 
the data have a cadence of a month for the first 
1.5 years, and 1 week for the rest of the 2.5 years in order to detect the 
shorter time lags effectively.  
In the second scenario (called ``Opt2''), the optical data are
obtained with a cadence of 2 weeks for the last 2.5 years. 

Because the mapping strategy for SPHEREx survey has not been determined yet, 
we conservatively assume that IR multi-epoch data will be obtained with 
seasonal gaps and a cadences of 6 days for 2 years. We adopt two  
observing scenarios, namely, ``IR1'' mode with a seasonal gap of 6 months and 
``IR2'' mode with a seasonal gap of 9 months. While IR1 mode may be applicable 
to the central part of the deep fields, the outer part can be more likely 
described by IR2 mode. By combining two scenarios in each observation, four 
different observing strategies are tested in this simulation (Table 1). 
In summary, we randomly generated 200 pairs of light curves at each 
position in a grid of four parameters (observed time lag, observing strategy, 
variability amplitude, and photometric error). 

In addition, the IR light curves are convolved with a top-hat transfer 
function to take into account the geometric effect of the dusty torus 
(\citealt{blandford_1982, almeyda_2020}). The width ($d \tau$) of the top-hat 
function is set to be a half of the time lag, while its maximum value is 
200 days (\citealt{lyu_2019}).

\begin{figure*}[h]
\centering
\includegraphics[width=160mm]{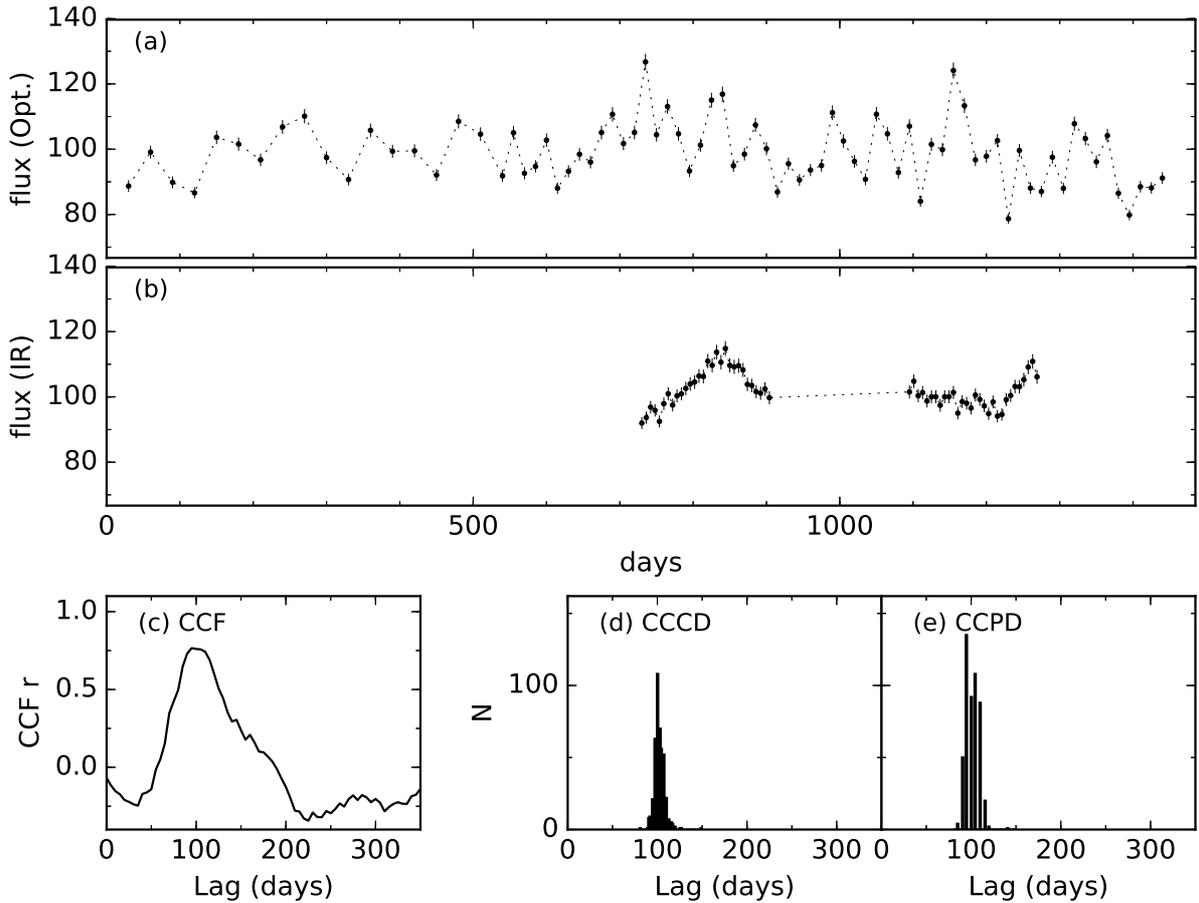}
\caption{Examples of simulated light curves and time lag measurements with 
ICCF. The variability amplitude is 10\%. 
The photometric error is set to be 2\% for both the optical and IR band.
The input time lag is $100$ days. The estimated time lag is 
$102.5^{+4.9}_{-4.9}$ days.
(a) Optical light curve obtained with ``Opt1''. 
(b) IR light curve is obtained with ``IR1''. 
(c) Cross correlation function (CCF), in which the cross-correlation Pearson 
coefficient ($r$) between the optical and IR light curves is shown as a 
function of a time lag. 
(d) Cross-correlation centroid distribution (CCCD) estimated using 
Monte-Carlo iterations. The centroid of the time lag is computed using the 
cross-correlation coefficients above 80\%\ of the peak value. We adopt 
the median of CCCD as a representative value of the time lag because it
is known to be less biased (\citealt{peterson_2014}).    
(e) Cross-correlation peak distribution (CCPD) estimated using Monte-Carlo 
iterations.
\label{fig:fig4}}
\end{figure*}

\begin{figure*}[h]
\centering
\includegraphics[width=160mm]{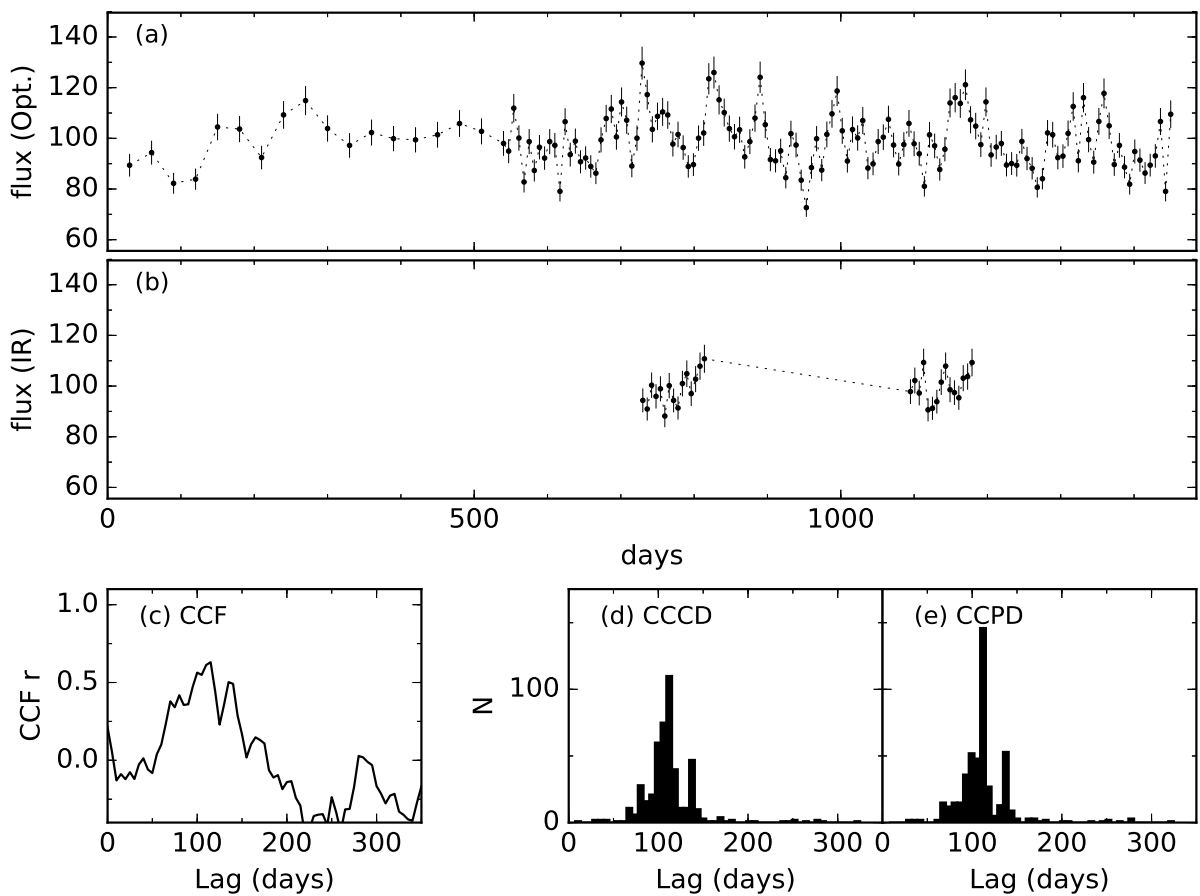}
\caption{Same as Figure 4, except that the photometric error is set to be 5\% 
for both the optical and IR band; optical light curve obtained with ``Opt2'';
IR light curve is obtained with ``IR2''. The estimated time lag is 
$110.1^{+27.4}_{-12.9}$ days.
\label{fig:fig5}}
\end{figure*}

\begin{figure*}[h]
\centering
\includegraphics[width=160mm]{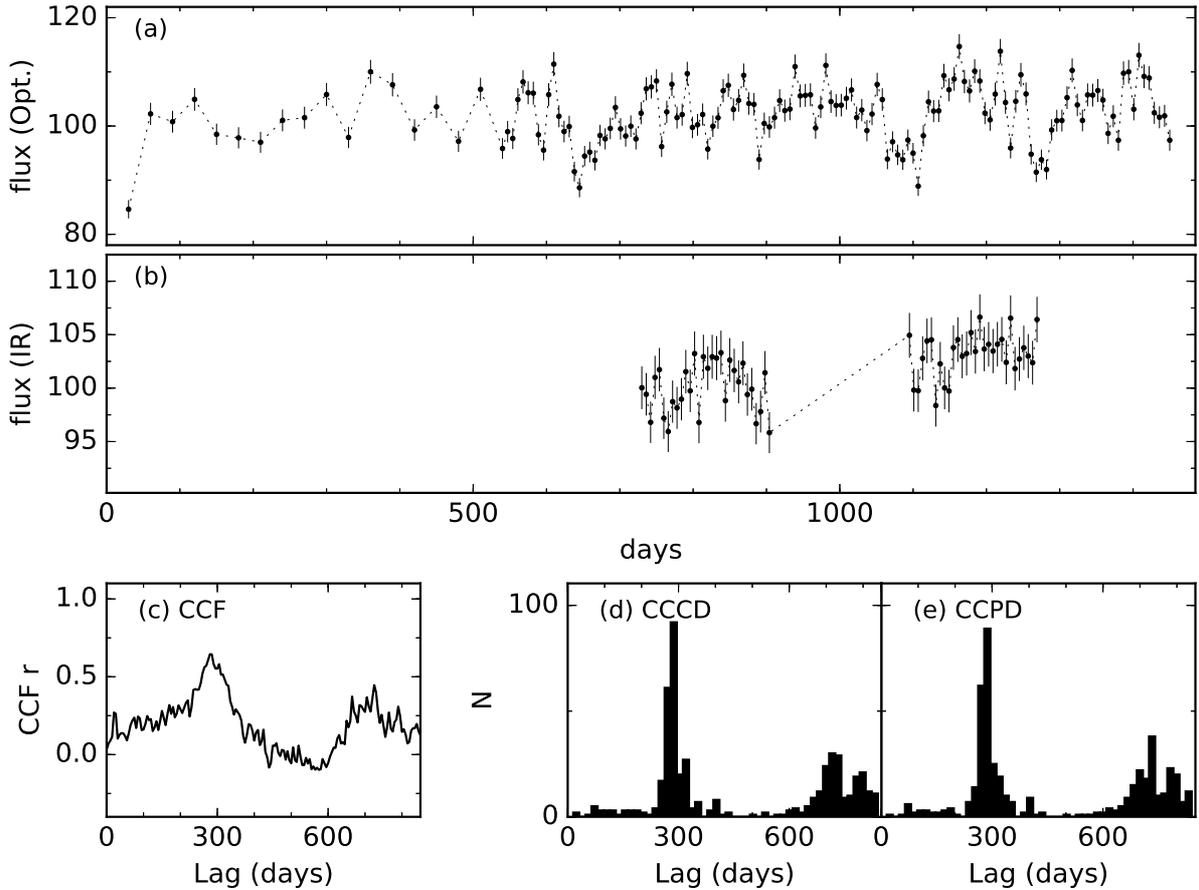}
\caption{Same as Figure 4, except that 
the variability amplitude is 5\%; optical light curve obtained with ``Opt2'';
the input time lag is $300$ days. 
The estimated time lag is $295.1^{+319.9}_{-45.4}$ days.
\label{fig:fig6}}
\end{figure*}

\begin{figure*}[h]
\centering
\includegraphics[width=160mm]{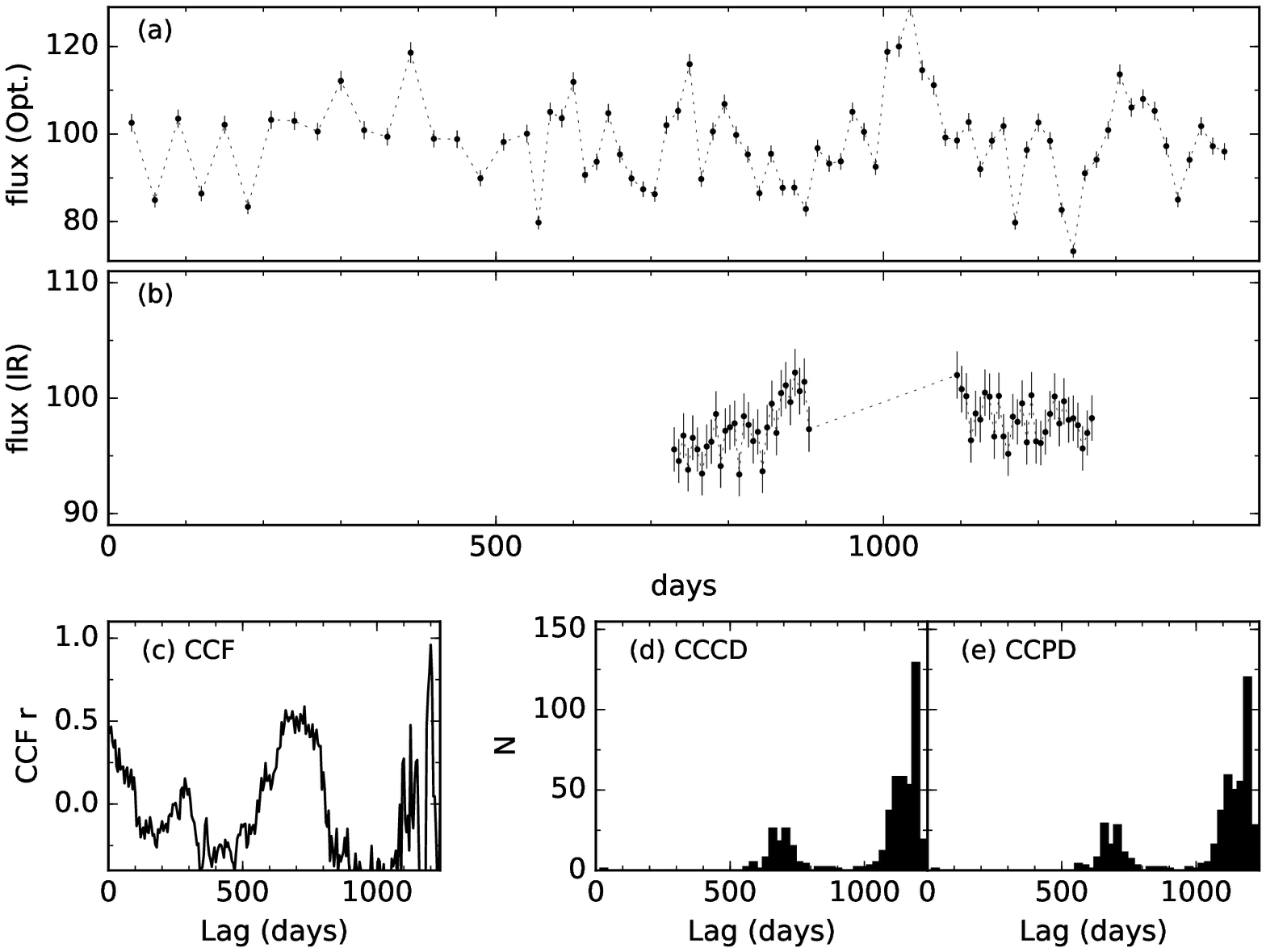}
\caption{Same as Figure 4, except that 
the input time lag is $700$ days. 
The estimated time lag is $1127.7^{+67.3}_{-414.8}$ days.
\label{fig:fig7}}
\end{figure*}

\section{Result}

To estimate the detection efficiency of time lags in the simulated light 
curves, 
we use a standard cross correlation technique (ICCF; \citealt{peterson_1998})
The uncertainty of the time lags is estimated by Monte Carlo simulation 
using randomly drawn subsets of photometric data from the simulated light 
curves (\citealt{gaskel_1987}). A python code PyCCF (\citealt{sun_2018}) is 
implemented for this calculation. 
 Prior to the time lag estimation, the optical light curves are 
convolved with the top-hat function, which is employed for the IR light 
curves.
We assume that the time lag is successfully recovered, if the 
estimated time lag is within 10\%\ of the input value. 
Examples of the time lag measurements along with the light curves are shown 
in Figure 4$--$7.

The simulation results are summarized in Table 1. 
In most cases, the detection rate for epoch1 (epoch2) is significantly 
larger than that for epoch3 (epoch4) by up to 40\%, revealing that the seasonal 
gaps in the IR observation are crucial in this experiment. However, we find 
little difference ($< 5\%$) between epoch1 (epoch3) and epoch2 (epoch4).
This finding indicates that the cadence in the optical monitoring data is less 
important. 

Figure 8 shows the detection efficiency of the time lags for 
the targets with 10\%\ of the variability amplitude. The detection rates 
are greater than 50\%\ for time lags of less than 300 days,
if the photometric error is smaller than 2\%. Larger time lags greater than
300 days is successfully recovered for less than half 
of the targets if the photometric error is larger than 2\%.
The detection rate dramatically decreases with increasing photometric error, 
revealing that a high S/N photometric data are essential for the precise
measurement of the time lag. 
Therefore, for the faint targets ($m_{3\mu m} > 19$ mag), S/N 
should be enhanced through the spectral binning and temporal binning. 
This requirement also needs to be taken into account for the optical 
observation.

Overall, for a variability amplitude of $\sim10$\%,
the detection efficiency for the time lags is relatively large
if the photometric error is less than 5\%.
By multiplying the time lag distribution of the primary targets to the 
interpolated success rate, it is expected to successfully estimate the time 
lags for $\sim 500$ objects if the observation strategy is properly designed 
(i.e., 5\%\ of the photometric error and 10\%\ of the variability amplitude).
If the high S/N optical monitoring data is available (i.e., 2\% of the 
photometric error), time lags can be successfully measured for
up to $\sim 900$ objects.

\begin{figure*}[h]
\centering
\includegraphics[width=150mm]{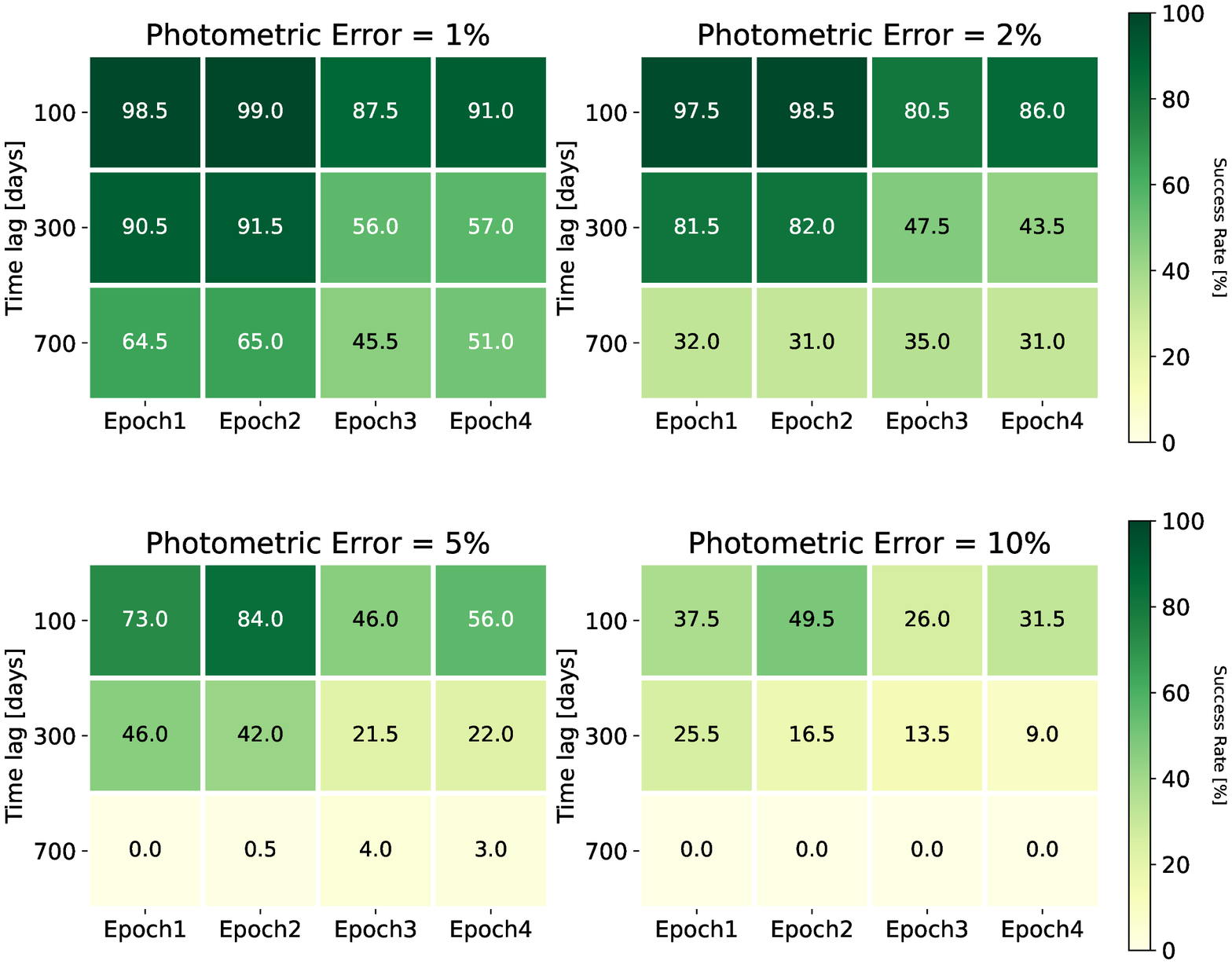}
\caption{Detection efficiency of time lags for a variability
amplitude of 10\%\ as a function of the input time lag and monitoring schedule. 
The detection rates are labelled in each bin.  
\label{fig:fig8}}
\end{figure*}

\begin{figure*}[h]
\centering
\includegraphics[width=150mm]{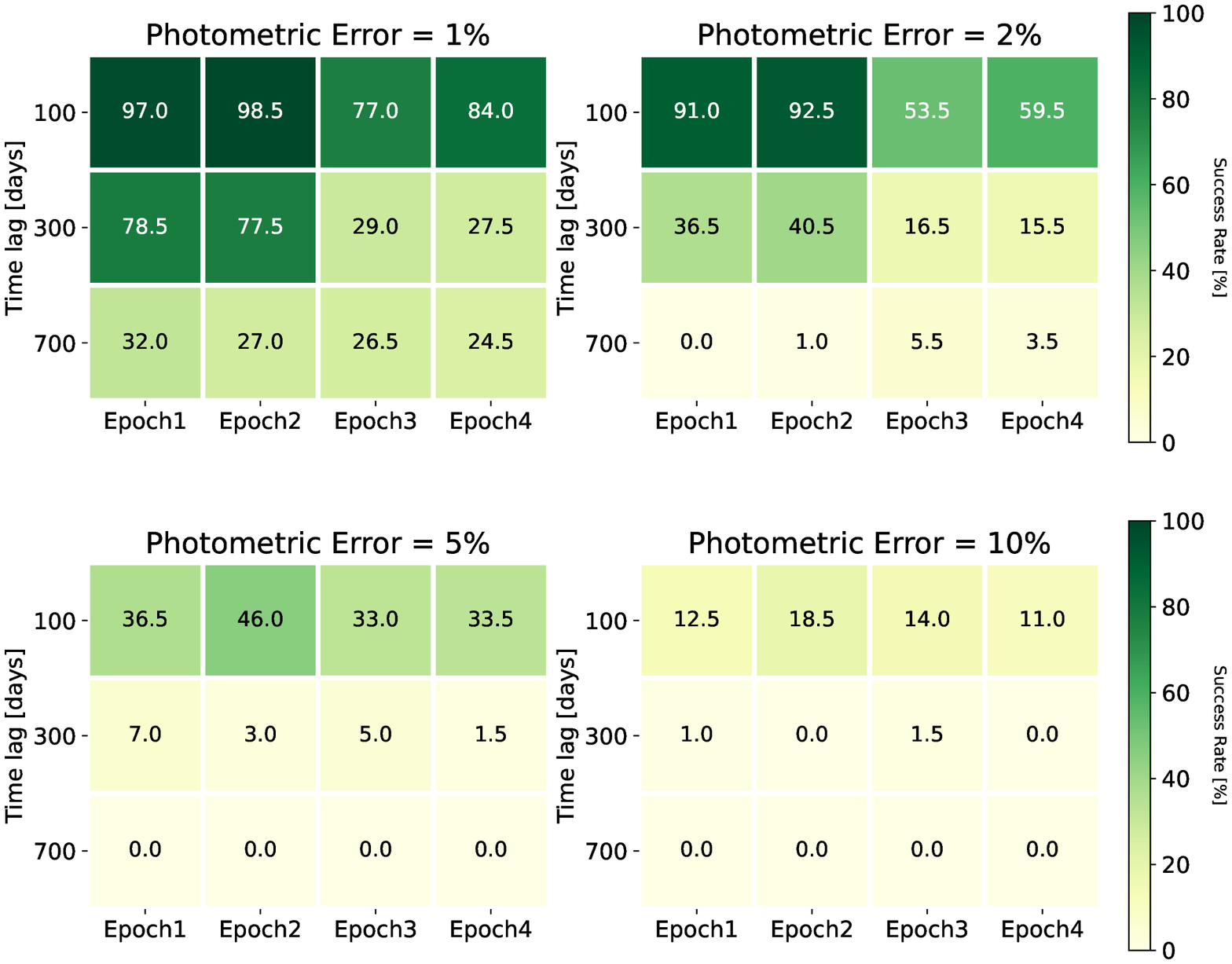}
\caption{Same as Figure 8, but for a variability amplitude of 5\%.
\label{fig:fig9}}
\end{figure*}


To take into account the smaller variability amplitude in IR than 
that in the optical band, and the light contamination from the host galaxy, 
we perform the same experiments under an assumption of a variability amplitude 
of 5\%. The results are shown in Figure 9. The success rate is significantly 
less than that obtained with a variability amplitude of 10\%. 
Not surprisingly, it is almost impossible to detect large 
time lags ($\sim 300$ days) if the photometric error is larger than 
the intrinsic variability amplitude (5\%).
However, in this worst case (5\%\ of the photometric error 
and 5\%\ of the intrinsic variability), we may still be able to recover the 
time lags for $\sim 200$ targets.

\section{Requirements for Complementary Optical Observation}
From the simulation, we find that the cadence for the early 
optical observation is less important, than the requirement for high S/N 
photometric data, which are the most crucial factor for the RM experiment. 
Through spectral and temporal binning, we can achieve S/N $>$ 10 
for 19 mag at 3$\mu m$. It approximately corresponds to 20 mag in the $B-$band. 
Therefore, to maximize the sample size, it is desirable that optical 
monitoring data reach up to $\sim21-22$ mag in the $B-$band. 
The existing/planned optical transient survey can be used for this 
purpose. For example, the Zwicky Transient Facility (ZTF) survey has
a limiting magnitude of $\sim 20.8$ mag in the $g-$band (\citealt{bellm_2019}),
which corresponds to $\sim 21.5$ mag in the $B-$band for typical QSOs
(\citealt{richards_2002,lupton_2005}).
Because its average cadence is $\sim3$ days, we can reach up to $\sim 22$ mag 
by combining it with multi-epoch data. Indeed, the Large Synoptic Survey 
Telescope (LSST) 
will provide the ideal optical dataset for the deep region around SEP with a 
limiting magnitude of $\sim 24.8$ mag in $g-$band with a single visit 
(\citealt{ivezic_2019}).  
If the public survey is unavailable, the optical monitoring observation 
should be executed using 1$-$2m class telescopes with
a wide field of view camera, such as the Korea Microlensing Telescope 
Network (KMTNet; \citealt{kim_2016}).  

With well-designed complementary optical observations (e.g., with 5\% 
of the photometric error at 20 mag\footnote{the median B-band magnitude 
of the primary target.}), one may be able to successfully detect 
time lags for $200-500$ 
objects depending on the intrinsic variability of the target QSOs. 
In addition, owing to the large sample size in the SPHEREx deep fields, one 
can estimate a composite lag using the stacking photometric data from 
individual targets (e.g., \citealt{fine_2012, li_2017}). 
On the other hand, previous studies on IR RM with a large sample 
relied on the WISE survey (\citealt{lyu_2019, yang_2020}), 
in which the IR data were obtained with a cadence of 6 months. 
Because the SPHEREx survey will be conducted with a much higher cadence 
(of about a few days) than WISE, the accuracy of the measurement for
the short time lags less than a couple of hundred days may be expected to 
be improved. 
The IR RM studies with the high cadence ground-based observations have been 
widely conducted for nearby AGNs (\citealt{koshida_2014, minezaki_2019}). 
The SPHEREx survey will extensively provide high S/N IR data for an unbiased 
large sample of AGNs at moderate redshifts, complementary to the previous 
studies.

\begin{table*}
\centering
\caption{
Success rate of RM simulations
\label{tab:jkastable1}}
\begin{tabular}{crcrrrr}
\toprule
Time lag & Variability & epoch & \multicolumn{4}{c}{Photometric Error} \\
\cline{4-7} 
 & & & 1\% & 2\% & 5\% & 10\% \\
(day) & & & (\%) & (\%) & (\%) & (\%) \\ 
(1) & (2) & (3) & (4) & (5) & (6) & (7) \\
\midrule
100 & 10\% & epoch1 & 98.5 &  97.5 &  73.0 &  37.5 \\
100 & 10\% & epoch2 & 99.0 &  98.5 &  84.0 &  49.5 \\
100 & 10\% & epoch3 & 87.5 &  80.5 &  46.0 &  26.0 \\
100 & 10\% & epoch4 & 91.0 &  86.0 &  56.0 &  31.5 \\
300 & 10\% & epoch1 & 90.5 &  81.5 &  46.0 &  25.5 \\
300 & 10\% & epoch2 & 91.5 &  82.0 &  42.0 &  16.5 \\
300 & 10\% & epoch3 & 56.0 &  47.5 &  21.5 &  13.5 \\
300 & 10\% & epoch4 & 57.0 &  43.5 &  22.0 &   9.0 \\
700 & 10\% & epoch1 & 64.5 &  32.0 &   0.0 &   0.0 \\
700 & 10\% & epoch2 & 65.0 &  31.0 &   0.5 &   0.0 \\
700 & 10\% & epoch3 & 45.5 &  35.0 &   4.0 &   0.0 \\
700 & 10\% & epoch4 & 51.0 &  31.0 &   3.0 &   0.0 \\
\\
100 & 5\% & epoch1 & 97.0 &  91.0 &  36.5 &  12.5 \\
100 & 5\% & epoch2 & 98.5 &  92.5 &  46.0 &  18.5 \\
100 & 5\% & epoch3 & 77.0 &  53.5 &  33.0 &  14.0 \\
100 & 5\% & epoch4 & 84.0 &  59.5 &  33.5 &  11.0 \\
300 & 5\% & epoch1 & 78.5 &  36.5 &   7.0 &   1.0 \\
300 & 5\% & epoch2 & 77.5 &  40.5 &   3.0 &   0.0 \\
300 & 5\% & epoch3 & 29.0 &  16.5 &   5.0 &   1.5 \\
300 & 5\% & epoch4 & 27.5 &  15.5 &   1.5 &   0.0 \\
700 & 5\% & epoch1 & 32.0 &   0.0 &   0.0 &   0.0 \\
700 & 5\% & epoch2 & 27.0 &   1.0 &   0.0 &   0.0 \\
700 & 5\% & epoch3 & 26.5 &   5.5 &   0.0 &   0.0 \\
700 & 5\% & epoch4 & 24.5 &   3.5 &   0.0 &   0.0 \\
\end{tabular}
\tabnote{
Col. (1): Input time lag.
Col. (2): Variability amplitude.
Col. (3): Name of the observing strategy.
Col. (4): Success rate for 1\%\ of the photometric error.
Col. (5): Success rate for 2\%\ of the photometric error.
Col. (6): Success rate for 5\%\ of the photometric error.
Col. (7): Success rate for 10\%\ of the photometric error.
}
\end{table*}

\section{Conclusion}
In order to test the feasibility of the RM experiments with SPHEREx dataset,
we search for the target QSOs in the SPHEREx deep regions and 
estimate the expected time lags of the sample using the size-luminosity
relations. We find that there are more than 1400 QSOs ($\tau_{\rm torus} 
\le 750$ days), which is suitable for the RM studies.
We perform the RM simulation to investigate the reliability of RM 
measurements. With the artificially generated light curves, we find
that the time lags can be successfully measured for $200-500$ objects
depending on the observation strategy (cadences and seasonal gaps) and 
photometric accuracy. 
In summary, in combination with the complementary optical observation, RM studies 
with SPHEREx can provide unique dataset to understand the physical properties of 
central structures of bright QSOs. 


\acknowledgments

We are grateful to anonymous reviewers for their critical
reviews that greatly helped to improve our manuscript.
This research was supported by Kyungpook National University Research Fund, 
2018.





\end{document}